\documentclass[twocolumn,nofootinbib,amsmath,amssymb]{revtex4}

\usepackage{graphicx}
\usepackage{dcolumn}
\usepackage{bm}
\usepackage{color}
\usepackage{cancel}
\usepackage{subfigure}

\newcommand{\Npix}{{N_{\rm pix}}}
\newcommand{\bhn}{{\bf \hat{n}}}
\newcommand{\deriv}{{\rm d}}
\newcommand{\bk}{{\bf k}}
\newcommand{\bhk}{{\bf \hat{k}}}
\newcommand{\bdtb}{\bar{\delta T}_b}
\newcommand{\dtb}{\delta T_b}
\def\max{{\rm max}}

\def\artanh{\mathrel{\rm arctanh}}

\newcommand{\msun}{M_{\odot}}
\newcommand{\bea}{\begin{eqnarray}}
\newcommand{\eea}{\end{eqnarray}}
\newcommand{\be}{\begin{equation}}
\newcommand{\ee}{\end{equation}}
\def\VEV#1{\left\langle #1 \right\rangle}

\begin{document}

\title{Galaxy-cluster masses via 21st-century measurements of
lensing of 21-cm fluctuations}

\author{Ely D. Kovetz$^{1}$ and Marc Kamionkowski$^{2}$}

\affiliation{$^1$Theory Group, Department of Physics and Texas
     Cosmology Center, The University of Texas at Austin, TX
     78712, USA}
\affiliation{$^2$Department of Physics and Astronomy, Johns
     Hopkins University, Baltimore, MD 21218, USA} 

\begin{abstract}
We discuss the prospects to measure galaxy-cluster properties
via weak lensing of 21-cm fluctuations from the dark ages and
the epoch of reionization (EOR). We choose as a figure of merit
the smallest cluster mass detectable through such measurements.
We construct the minimum-variance quadratic estimator for the
cluster mass based on lensing of 21-cm fluctuations at multiple
redshifts.  We discuss the tradeoff between frequency bandwidth,
angular resolution, and number of redshift shells available for
a fixed noise level for the radio detectors.  Observations of
lensing of the 21-cm background from the dark ages will be
capable of detecting $M\gtrsim10^{12}\, h^{-1}M_{\odot}$ mass
halos, but will require futuristic experiments to overcome the
contaminating sources. Next-generation radio measurements of
21-cm fluctuations from the EOR will, however, have the
sensitivity to detect galaxy clusters with halo masses
$M\gtrsim10^{13}\, h^{-1}M_{\odot}$, given enough observation
time (for the relevant sky patch) and collecting area to maximize their resolution
capabilities.
\end{abstract}


\maketitle

\section{Introduction}

The hyperfine transition of neutral hydrogen at 21 cm
provides a unique source of cosmological information from the
epoch of reionization (EOR) and the dark ages
\cite{Furlanetto,Pritchard:2011xb}.  It is the target of 
several ongoing and near-future ground-based experiments
\cite{LOFARSKA}, as well as more distant prospects such as a
Lunar-based observatory \cite{Lunar}. During the two
relevant cosmological epochs, the dark ages and the EOR, the
hyperfine transition is observed in absorption or emission,
respectively, against the cosmic microwave background (CMB).  By
measuring this signal at different frequencies, the large
redshift volume of these two epochs can be used to generate
independent images of the spatial distribution of neutral
hydrogen at different redshifts
\cite{Zaldarriaga,LoebZald,CAMB_sources}.

The image of the 21-cm signal from the dark ages and/or EOR
should appear on the sky as a random field described by some
statistically isotropic two-point correlation function. 
If, however, that image is distorted by weak gravitational
lensing from foreground matter---either the large-scale
inhomogeneous distribution of mass in the Universe or by
discrete objects, like galaxy clusters---then there may be local
departures from statistical isotropy induced
\cite{ZahnZald,Metcalf,MetcalfWhite,Hilbert,LuPen,Sigurdson:2005cp,Book:2011dz}.
Measurement of these local departures from statistical isotropy
may thus allow a measurement of the distribution of this
intervening matter.

The effects of gravitational lensing of 21-cm anisotropies are the
same as the analogous effects on CMB fluctuations and can
therefore be analyzed with the tools developed for lensing of
the CMB \cite{Seljak,Okamoto,HuDeDeo}.  However, the extension
of 21-cm fluctuations to far smaller angular scales (limited in
principle only by the baryonic Jeans mass \cite{LoebZald}) than
CMB fluctuations (which are suppressed on small scales by Silk
damping \cite{Silk:1967kq}), and the possibility to see
images of the 21-cm background at multiple redshifts, make
21-cm lensing far more promising, ultimately, for weak-lensing
studies. 
In particular, constraining the parameters of galaxy cluster mass profiles using weak lensing reconstruction can be a useful tool in probing the evolution of dark energy \cite{Albrecht} and studying the properties of dark matter (e.g. through the characterization of substructure or the mapping of its distribution in merging clusters). 

High-resolution imaging of the mass distribution in
galaxy clusters using 21-cm lensing was investigated in
Ref.~\cite{MetcalfWhite} (and with simulations in
Ref.~\cite{Hilbert}).  This work concluded that while
forthcoming experiments may have the potential to provide some
initial detections of galaxy clusters, the full promise of the
technique will likely have to await subsequent generations.

The goal of this paper is to revisit lensing of the 21-cm
background by galaxy clusters  with an analytic treatment
aimed primarily to help understand the dependence of the
detectability of the signal on experimental parameters.  The
aim will be to clarify the experimental requirements for such
detections and to assist in the design of experiments to make
such measurements.  More specifically, we use as a figure of
merit the smallest galaxy-cluster mass detectable by a given
experimental configuration and then investigate the dependence
of this threshold mass on the experimental configuration.

The plan of the paper is as follows:  In
Section II, we review the 21-cm signal and examine its dependence on the observation frequency
and on the bandwidth over which the signal is observed.  In Section III we 
discuss the noise power spectrum of radio interferometers and
study its dependence on frequency, bandwidth, angular resolution,
observation time, and collecting area.  In Section IV we review
how lensing of the 21-cm background by galaxy-cluster masses is
accomplished.  We present
a quadratic estimator for the weak-lensing convergence and
derive the noise with which it can be measured. In Section V we
construct a minimum-variance estimator for the galaxy-cluster
mass obtained from the lensing convergence and derive a formula for the smallest detectable galaxy-cluster mass as a function of the various experimental parameters, assuming an Navarro-Frenk-White (NFW) mass profile. 
In Section VI we show the results for galaxy-cluster mass detectability with next generation interferometers as well as futuristic ideal experiments and discuss different tradeoffs between the experimental parameters.
In Section VII we compare the prospects for
mass measurements from lensing of 21-cm fluctuations with other
mass measurements, mention possible improvements to our estimator and
discuss some relevant subtleties. We conclude in Section VIII.

\section{21-cm emission/absorption signal}

\begin{figure}[b!]
\includegraphics[width=0.8\linewidth]{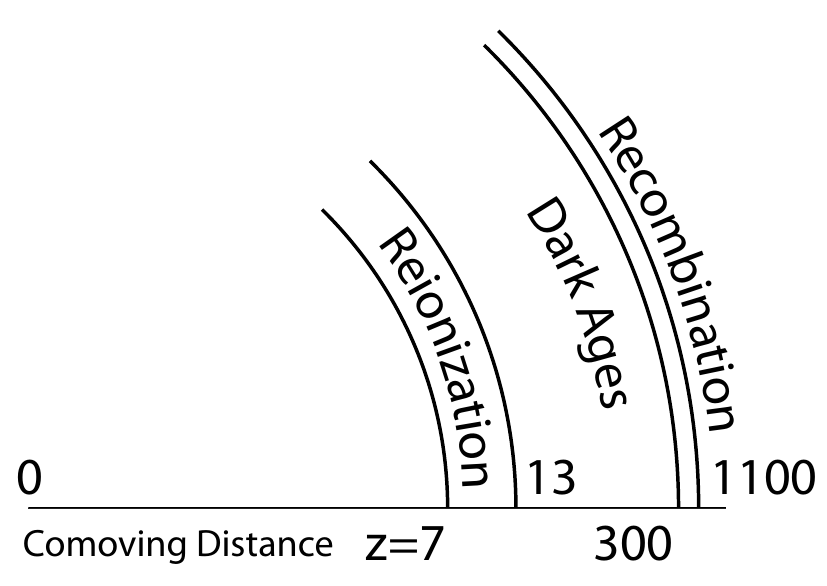}
\caption{An illustration of the comoving volume of the universe demonstrating the huge potential for extracting valuable information from 21-cm radiation in the dark ages and the reionization epochs.}
\label{fig:Volume}
\end{figure}

We begin by reviewing the physics responsible for producing the
21-cm signal from the dark ages and the EOR, whose relative comoving volumes are illustrated in Fig. \ref{fig:Volume}.
Defining the spin temperature $T_s$ as the excitation
temperature of the hyperfine transition (characterizing the
ratio between the number densities of hydrogen atoms in the
excited and ground-state levels), the rest-frame brightness
temperature of a patch of the sky is given by $T_b=T_{\rm
CMB}e^{-\tau}+T_S(1-e^{-\tau})$, where the optical depth for the
hyperfine transition is
\cite{field59}
\bea
\tau & = & \frac{ 3 c^3 \hbar A_{10} \, n_{\rm HI}}{16 
k \nu_0^2 \, T_S \, H(z) },
\label{tau} 
\eea
$A_{10} = 2.85 \times 10^{-15}\, {\rm s}^{-1}$ is the Einstein
coefficient for the transition, $\nu_0=1420 \, {\rm MHz}$ its
rest-frame frequency and $n_{\rm HI}$ is the local neutral
hydrogen density.  The brightness temperature in this patch, at an
observed frequency $\nu$ corresponding to a redshift $1+z=\nu_0/\nu$,
and the CMB is given by the difference
\bea
\delta T_b(\nu) & \approx & \frac{T_S - T_{\rm CMB}}{1+z} \, \tau .
\label{Tbright}
\eea
Hence if the excitation temperature $T_s$ in a region differs from
that of the CMB, the region will appear in emission (if $T_s>T_{\rm CMB}$)
or absorption (if $T_s<T_{\rm CMB}$) against the CMB. 

During the dark ages, there is a redshift period $30<z<200$
where neutral hydrogen should be visible in absorption against
the CMB, as the spin temperature is coupled to the gas
temperature [and cools adiabatically as $(1+z)^2$] while
collisions are efficient and drops below the CMB temperature
[which only cools as $(1+z)$]. This process peaks at
$z\sim70$ and lasts until the Hubble expansion renders
collisions inefficient and $T_s\sim T_{\rm CMB}$ again at
$z\sim30$. The signal from this early epoch is not affected by
nonlinear density structures nor contaminated by astrophysical
sources, which have yet to form. 

A 21-cm signal is also accessible during the EOR, at redshifts $7 \lesssim z \lesssim 13$, when
newly formed structure heated up neutral hydrogen but 
before the hydrogen became fully ionized.  Predicting the
brightness-temperature signal in this epoch is much harder, as
astrophysical noise sources are substantial and our
uncertainties as to the beginning and duration of this period
are significant. Unlike the dark ages signal, in order to plot the signal during
reionization we need to trace the redshift behavior of the
neutral gas fraction which determines the optical depth,
Eq.~(\ref{tau}), for the hyperfine transition.

\begin{figure}[htbp]
\includegraphics[width=0.8\linewidth]{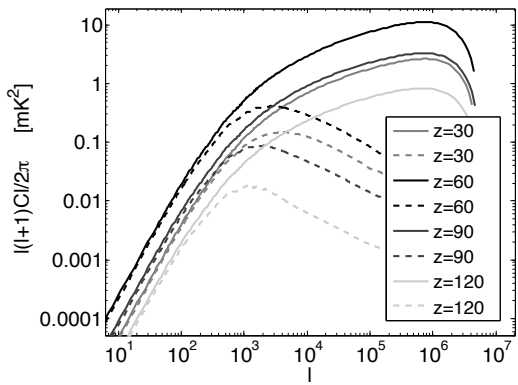}
\caption{21-cm power spectrum for different redshifts during the
dark ages, calculated using CAMB Sources
\cite{CAMB_sources}. Solid lines are calculated in the
narrow-bandwidth limit and the signal is damped by the effect of
baryon pressure at $l\gtrsim10^6$. Dashed lines are for
$\Delta\nu=0.1\,{\rm MHz}$, where the signal is damped due to
line-of-sight averaging over the bandwidth above $l\gtrsim10^3$
as predicted by Eq.~(\ref{eq:damping}) for these redshift ranges.} 
\label{fig:CLDarkAges}
\end{figure}

In the Appendix, we review the power spectrum of intensity fluctuations of 21-cm radiation. Eqs.~(\ref{eq:21PowerSpectrum})-(\ref{eq:21PowerSpectrumSmallScales}) describe the angular power spectrum and its approximated form in the limit of large and small scales. In observations of these fluctuations, an important factor is the damping that results from line-of-sight averaging in a width $\delta r$ (corresponding to an observed bandwidth $\Delta\nu$) around the distance $r$ to the desired frequency $\nu$, whose scale is determined for a given radio interferometer through the relation
\cite{Zaldarriaga,LoebZald}
\be
\delta r/r\simeq 0.5(\Delta\nu/\nu)(1 + z)^{-1/2}.
\label{eq:damping}
\ee 
In Fig.~\ref{fig:CLDarkAges} we plot the 21-cm dark ages signal power
spectrum, Eq.~(\ref{eq:21PowerSpectrum}), in several redshifts using
CAMB \cite{CAMB_sources} in the limit of sharp frequency bandwidth
and with a bandwidth of $\Delta\nu=0.1\,{\rm MHz}$ (resulting in damping above $l\sim10^3$) in the linear
regime.
The signal peaks at redshift $z\sim60$ where the deviation between the spin temperature and that of the CMB is maximal \cite{LoebZald}. We see that for a given bandwidth, the signal is roughly within the same order of magnitude for most redshifts up to very small angular scales, corresponding to arcsecond resolutions. 

\begin{figure}[htbp]
\includegraphics[width=0.8\linewidth]{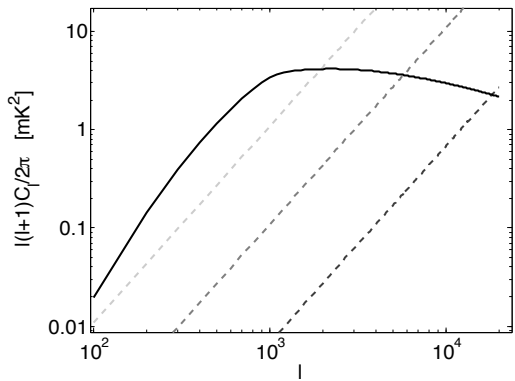}
\caption{21-cm EOR power spectrum at redshift $z\!=\!7$, assuming complete and instantaneous reionization just after $z\!=\!7$, using Eqs.~(\ref{eq:21PowerSpectrumLargeScales}) and (\ref{eq:21PowerSpectrumSmallScales}) (with an amplitude $\bdtb(\nu)\sim20\,{\rm mK}$ \cite{Zaldarriaga}). The solid line is the signal power spectrum for a bandwidth of $\Delta\nu=1~{\rm MHz}$, which is damped around $l\lesssim10^3$ for this redshift according to Eq.~(\ref{eq:damping}). Dashed lines are noise power spectra for SKA (light gray), SKA with $10\times t_0$ (medium gray) and SKA with $10\times t_0$ and $4\times f_{\rm cover}$ (dark gray). We see that with the final setup we can reach the maximum $l$ observable with SKA's baseline of $6~{\rm km}$ which is roughly $l_{\rm cover}\sim2\times10^4$.}
\label{fig:CLEoR}
\end{figure}

In Fig.~\ref{fig:CLEoR} we plot the 21-cm EOR signal power spectrum, assuming reionization is instantaneous and complete at $z\!=\!7$, using the approximations in Eqs.~(\ref{eq:21PowerSpectrumLargeScales}) and (\ref{eq:21PowerSpectrumSmallScales}) for a bandwidth of $1\,{\rm MHz}$. The damping scale is again approximately $l\sim10^3$ (as the redshift of reionization is an order of magnitude smaller), and the signal is roughly constant for small scales up to $\gtrsim10^4$, corresponding to slightly better than arcminute resolutions.

In the next section we compare the 21-cm signal with the noise power
spectrum estimation of radio interferometers, before heading on
to discuss the reconstruction of galaxy clusters from the weak
lensing signal of the 21-cm radiation.

\section{Experimental Prospects}

Following Ref.~\cite{Zaldarriaga}, we examine the noise power
spectrum of a radio interferometer array. 
We denote by $l_{\rm cover}(\nu)=2\pi D/\lambda$ the maximum
mode at frequency $\nu$ (corresponding to wavelength $\lambda$)
that can be measured with an array of dishes with  maximum
baseline $D$ covering a total area $A_{\rm total}$ with a covering
fraction $f_{\rm cover}\equiv N_{\rm dish} A_{\rm dish} /A_{\rm
total}$. 
For uniform Fourier area coverage, a system temperature $T_{\rm
sys}$, a frequency window $\Delta \nu$, and an observing time
$t_o$, the noise power spectrum is 
\be
l^2 C_l^n= {(2\pi)^3 T_{\rm sys}^2(\nu) \over \Delta \nu t_o  f_{\rm cover}^2} \ \left({ l\over l_{\rm cover}(\nu)}\right)^2. 
\ee
Thus, for uniform Fourier coverage in an experiment targeting a
frequency $\nu$ with bandwidth $\Delta\nu$, we conclude that
$C^n_l \sim {\rm const}\equiv 2\pi\beta(\nu)(f_{\rm cover}^2\,
t_0 \Delta\nu)^{-1}$, where $\beta(\nu)= (2\pi)^2 T_{\rm
sys}^2(\nu) [l_{\rm cover}(\nu)^2]^{-1}$. As we can see from
Figs.~\ref{fig:CLDarkAges} and \ref{fig:CLEoR}, on the relevant (small) scales for cluster reconstruction, the 21-cm radiation
power spectrum at a given frequency and bandwidth behaves roughly (up to an order of magnitude) as $l^2C_l \sim {\rm const}
\equiv 2\pi\alpha(\nu,\Delta\nu)$. To find the value $l_{\rm
max}$ beyond which the noise $C^n_l$ is no longer negligible, we
compare $C_{l_{\rm max}}=C^n_{l_{\rm max}}$ to find
\be
l^2_{\rm max}=\frac{\alpha(\nu,\Delta\nu)}{\beta(\nu)} (f_{\rm cover}^2\, t_0 \Delta\nu).
\label{l_max}
\ee
This quantity depends trivially on the fractional coverage and observation time (they only affect the noise). The dependence on frequency and bandwidth is more elaborate, as we discuss later on below.

Observing the 21-cm anisotropies during the dark ages will be very challenging.
The redshift range $30<z<200$ of absorption of the CMB at 21-cm
($\nu = 1420\, {\rm MHz}$) in the rest frame of the
neutral-hydrogen gas corresponds to very low frequencies $7\,
{\rm MHz} < \nu < 46 \,{\rm MHz}$, where the sky temperature (dominated by
foreground sources of synchrotron emission from the Galaxy and
from extragalactic sources) in regions of minimum emission at
high Galactic latitudes, approximately given by $T_{\rm sys}
\sim 180\left(\nu/180\,{\rm MHz}\right)^{-2.6}$~K, reaches
$\sim 10^4-10^6 \,{\rm K}$, many orders of magnitude above the
signal $\sim 1\, {\rm mK}$ (as seen in Figs.~\ref{fig:CLDarkAges} and \ref{fig:CLEoR}).

While both Galactic and extragalactic emissions vary smoothly
and could be possibly subtracted by taking observations at two
closely spaced frequencies, the additional sources of
interference are harder to overcome. Terrestrial radio frequency
interference is abundant in this range and is not spectrally
smooth. Another source of contamination is the ionosphere which
causes phase distortions in the cosmic signal and turns opaque
at frequencies $\nu \lesssim 20\, {\rm MHz}$ (corresponding to
$z > 70$).
A futuristic experiment, based on placing a dark ages
observatory on the far side of the Moon (which has no permanent
ionosphere and its far side is shielded from terrestrial radio
interference) was suggested in Ref.~\cite{Lunar}. With a
baseline on the order of $\gtrsim10-100\,{\rm km}$, angular
resolutions corresponding to $l_{\rm max}\sim10^4\!-\!10^5$ (for
source redshifts $z=30-300$) might be reached in such an
experiment, compensating for covering fraction and system
temperature with prolonged observation time. With sufficient
frequency coverage, the huge volume of CMB absorption in neutral
hydrogen during the dark ages, $30 \lesssim z \lesssim 200$, can
be used to beat down the noise (by combining redshift slices, as we
describe in the next Section) and measure the mass profile
parameters of galaxy clusters to high accuracy.

To reach a measurement within the decade, the best candidate is
the currently planned next generation experiment to measure
21-cm emission from neutral hydrogen during the EOR, the
\emph{Square Kilometer
Array}\footnote{http://www.skatelescope.org} (SKA). With a
design based on an extended region of $D\! \sim\! 6$ km where
$l_{\rm cover}(\nu)\!\sim\! 10^4$ for the relevant frequencies in
the epoch of reionization and $f_{\rm cover}\! \sim\! 0.02$,
SKA can reach an observing time of order $t_0\sim\!1000\, {\rm
hrs}$ in one season covering up to an area of $\sim2\pi\, {\rm
sr}$ in the sky.

In Fig.~\ref{fig:CLEoR} we plot several noise power spectra matching the current plans for a three-season run of SKA measuring at a frequency corresponding to redshift $z\!=\!7$ with a bandwidth of $1\,{\rm MHz}$, as well as scenarios with an order of magnitude larger observation time, and with four times the coverage fraction. With the current plan, the maximum scale observable with a signal to noise greater than one is only $O(l_{\rm max}\!\sim\!10^3)$. To reach $l_{\rm max}\!\gtrsim\!10^4$ will require a considerable increase in coverage fraction and/or observation time.

\section{Lensing of 21-cm Fluctuations by a Galaxy Cluster}

We now review the distortion to the 21-cm fluctuations induced
by weak gravitational lensing.  We assume that the lensing
distortion takes place over a relatively small region of the
sky, as should occur for lensing by a galaxy cluster, so that we
can work in the flat-sky limit, where the analytic expressions
are simpler.

Let $I_0(\vec\theta)$ be the 21-cm intensity at position
$\vec\theta$ on the sky.  Lensing will deflect photons from
$\vec\theta$ by an amount $\delta\vec\theta(\vec\theta) = \nabla
\phi(\vec\theta)$.  Here, $\phi(\vec\theta)$ is the projected
potential, related to the convergence $\kappa(\vec\theta) = 
\Sigma(\vec\theta)/\Sigma_{\rm cr}$, by $\nabla^2\phi = 2 \kappa$,
where $\Sigma(\vec\theta)$ is the surface mass density of the
intervening cluster, and $\Sigma_{\rm cr}^{-1}=4 \pi G D_d
D_{ds}/(c^2 D_{s})$ is the critical surface mass density in
terms of the observer-lens, lens-source, and observer-source
angular-diameter distances $D_d$, $D_{ds}$, and $D_{s}$,
respectively.  The observed intensity is thus $I(\vec\theta) =
I_0(\vec\theta+\delta\vec\theta) \simeq I_0(\vec\theta)+
\delta\vec\theta \cdot \nabla I_0(\vec\theta)$.

We suppose that intensity is measured over some square
patch of sky of solid angle $\Omega$, decomposed into $N_{\rm
pix}$ pixels at positions $\theta_i$, for $i=1,2,\ldots,\Npix$,
surrounding the cluster.  The intensity can then
be written in terms of Fourier coefficients,
\begin{equation}
     I_{\vec l} = \frac{\Omega}{\Npix} \sum_{\vec\theta_i} e^{i
     \vec l \cdot \vec\theta_i} I(\vec\theta_i),
\end{equation}
as
\begin{equation}
     I(\vec\theta_i) = \frac{1}{\Omega} \sum_{\vec l} e^{-i \vec l
     \cdot \vec\theta}I_{\vec l} ,
\end{equation}
where the sum is over the $\Npix$ Fourier modes $\vec l$.  The
two-point intensity correlations are described by a power
spectrum $C_l$ defined by
\begin{equation}
     \VEV{ I_{\vec l} I_{\vec l'}^*} = \Omega C_l \delta_{\vec
     l,\vec l'}
\end{equation}
The Fourier coefficients for the observed intensity are related
to those of the unlensed intensity and deflection field by
\begin{equation}
     I_{\vec l} = I_{0 \vec l} - \frac{1}{\Omega} \sum_{\vec l'}
     \vec l' \cdot (\vec l- \vec l') \phi_{\vec l'} I_{0\, \vec
     l-\vec l'},
\end{equation}
from which it follows that for a fixed deflection field, the observed intensity satisfies 
\begin{equation}
     \VEV{ I_{\vec l_1} I_{\vec l_2}} = \Omega C_{l_1}\delta_{\vec
     l_1,-\vec l_2} + \phi_{\vec L} 
     \vec L \cdot (\vec l_1
     C_{l_1} + \vec l_2 C_{l_2} ),
\end{equation}
where $\vec L = \vec l_1 +\vec l_2$ and the ensemble average is for a fixed deflection field.

Since $\kappa_{\vec L} = -L^2 \phi_{\vec L}/2$, 
each $\vec l_1$-$\vec l_2$ pair of
measured intensity Fourier modes with wavevectors $l_1+l_2 =
\vec L$ provides an estimator,
\begin{equation}
     \widehat{ \kappa_{\vec L}^{\vec l_1, \vec l_2}} =
     -\frac{L^2}{2} \frac{
     I_{\vec l_1} I_{\vec l_2}}{      \vec L \cdot (\vec l_1
     C_{l_1} + \vec l_2 C_{l_2} )},
\end{equation}
and the variance of this estimator is
\begin{equation}
     \VEV{ \left( \widehat{ \kappa_{\vec L}^{\vec l_1, \vec
     l_2}} \right)^2} =  \frac{L^4}{4} \frac{C_{l_1}^{\rm
     map} C_{l_2}^{\rm
     map} \Omega^2}{ \left[ \vec L \cdot (\vec l_1
     C_{l_1} + \vec l_2 C_{l_2} ) \right]^2},
\end{equation}
where $C_l^{\rm map}=C_l+ C_l^{\rm n}$ is the power spectrum of
the intensity map, including the noise $C_l^{\rm n}$.

We can then sum the estimators $\widehat {\kappa_{\vec L}^{\vec
l_1,\vec l_2}}$ over all $\vec l_1+\vec l_2 = \vec L$ pairs (correcting for double counting of triangles with $\vec l \leftrightarrow \vec L-\vec l$) with
inverse-variance weighting to obtain the minimum-variance
estimator, 
\begin{eqnarray}
     \widehat{\kappa_{\vec L}} 
     & = & -\frac{\Omega N_{\vec{L}}}{L^2 } 
      \sum_{\vec l} \frac{I_{\vec l} I_{\vec L
     -\vec l} }{ \vec L \cdot(\vec l C_l + (\vec L-\vec l)
     C_{|\vec L-\vec l|})} \nonumber \\
     & & \times 
     \left[ \frac{C_l^{\rm map} C_{|\vec
     L -\vec l|}^{\rm map} \Omega^2}{ [\vec L \cdot(\vec l C_l + (\vec L-\vec l)
     C_{|\vec L-\vec l|})]^2}\right]^{-1},
\end{eqnarray} 
and 
\begin{equation}
     N_{\vec{L}}^{-1} = \frac{2\Omega}{L^4}      \sum_{\vec l} 
     \frac {
     [\vec L \cdot(\vec l C_l + (\vec L-\vec l)   C_{|\vec
     L-\vec l|})]^2}
     {C_l^{\rm map} C_{|\vec L -\vec l|}^{\rm map}
     \Omega^2}
\label{eq:huestimator}
\end{equation}
is the noise power spectrum for $\kappa_{\vec L}$.
Equivalently, $\langle| \widehat{\kappa_{\vec L}} |^2\rangle=(2\pi)^2\delta(0)N_{\vec{L}}=\Omega N_{\vec{L}}$ is the variance with which
$\kappa_{\vec L}$ can be measured.

We now use the continuum limit $\Omega^{-1} \sum_{\vec l} \to
\int d^2l/(2\pi)^2$ and employ the simplifying assumption,
Eq.~(\ref{l_max}), that the noise is small,
$C_l^{\rm n} \ll C_l$, so that $C_l^{\rm map}\simeq C_l$, up to
some scale $l_\max$. 
Then, in the limit $L \ll l$ we approximate $|\vec L-\vec l|
\simeq l-L\cos\phi$, where $\cos\phi\equiv \hat L\cdot \hat l$,
and then to first order, $C_{|\vec L-\vec l|} \simeq C_l - L
(\cos\phi) (\partial C_l/\partial l)$, which
yields \cite{ZahnZald}
\begin{eqnarray}
     N_{\vec{L}}^{-1} \!&\simeq&\!  \frac{4}{L^4}\int {\frac{d^2l}{(2\pi)^2}}
     \frac{\left[\vec L\cdot \vec l C_{l} +\vec L\cdot (\vec L-\vec l) C_{|\vec L-\vec l|}\right]^2}{2 \,
     C_l \, C_{|\vec L-\vec l|}} \nonumber \\
                &\simeq&\!
                \frac{l_\max^2}{2\pi}\left[1+
                \frac{\partial\ln C_l}{\partial \ln l}
                  + \frac{3}{8} \left(\frac{\partial\ln
                  C_l}{\partial \ln l}\right)^2\right].
\label{eqn:vareqn}
\end{eqnarray}
Using the approximation leading to Eq.~(\ref{l_max}), that the the noise for a
single slice in an experiment at frequency $\nu$ with bandwidth
$\Delta\nu$, a maximum baseline corresponding to $l_{\rm
cover}$, a coverage fraction $f_{\rm cover}$, and observation
time $t_0$,  is approximately,
\be
N_{\vec{L}} \sim \frac{4\pi}{ l_\max^2}=\frac{4\pi\beta(\nu)}{\alpha(\nu,\Delta\nu)(f_{\rm cover}^2\, t_0\Delta\nu)}.
\label{VarLimit}
\ee
  
We can increase our signal considerably by changing the
frequency at which the 21-cm map is made and thereby focus on
spherical shells of neutral hydrogen at different redshifts. The
above results for a single redshift slice can be extended to
make use of the full redshift volume \cite{ZahnZald} by
discretizing the $z$ direction into radial components so that
$C_{l,k}$ is the power in a mode with angular component $l$ and
radial components $k=2\pi j/ R$, where $R$ is the
total radial length of the volume. Under the assumption that
modes with different $k$ are independent, a total-volume
estimator is built by summing the individual estimators, and the
corresponding noise variance is
\bea
\frac{L^4/4}{N_{\vec{L}}}=\sum\limits_k \int {\frac{d^2l}{(2\pi)^2}} \frac{[\vec L\cdot \vec l C_{l,k}
    +  \vec L\cdot (\vec L-\vec l) C_{|\vec L-\vec l|,k}]^2}{2 C^{\rm
    map}_{l,k}C^{\rm map}_{|\vec L-\vec l|,k}} \nonumber \\
  \label{volumenoise}
\eea

In general, to estimate the number of slices
available, we can think of 21-cm maps at two
different frequencies that correspond to spherical shells
separated along the line of sight by a comoving distance $\Delta
r$. These will be statistically independent at the highest $l$
provided that $(\Delta r/r_{\nu}) \gtrsim l^{-1}$.  An
experiment that covers a spatial range $\delta r$ or a frequency
range $\nu_1-\nu_2$ around a frequency $\nu$ will yield a total
number of $N_z\sim(\delta r/\Delta r) \simeq l(\delta
r/r_{\nu})\simeq 0.5l((\nu_1-\nu_2)/\nu)(1 + z)^{-1/2}$
statistically independent maps, which is roughly $N_z\sim1500$
for the frequencies corresponding to the EOR at the maximum resolution of SKA and up to $N_z\sim5000$ for a
dark ages observatory with a baseline of $100\,{\rm km}$ (here we
neglect the fact that different frequency bins are also somewhat
correlated by contaminations). In practice, however, the
tradeoff is a complicated one. While the signal is larger for
narrow bandwidths, we saw in the last Section that the noise
power spectrum around a given frequency increases as the
frequency bandwidth is decreased, which limits the actual number
of slices that can be combined in Eq.~(\ref{volumenoise}).

\section{Estimating Cluster Properties}

Applying the estimator in Eq.~(\ref{eq:huestimator}) for the
convergence $\widehat{\kappa}_L$ to a patch of sky around a galaxy
cluster (assuming we know the location of its center to enough
accuracy, say from other sources such as Sunyaev-Zeldovich surveys), we can
retrieve a 2D image of the weak-lensing convergence of the
cluster which we can study versus theory or simulation.  
The total number of pixels in Fourier space is the same as in real space.  We then denote the $N$ Fourier wavenumbers as $\vec{L_i}$ for $i=1,2,\ldots,N$. If $\widehat{\kappa_{\vec{L}_i}}$ is the
measured value for the pixel $i$ with variance
$\langle|\widehat{\kappa_{\vec{L}_i}}|^2\rangle=\Omega N_{\vec{L}_i}$
and the corresponding theoretical value is $\kappa^{\rm
th}_{\vec{L}_i}(M_{\rm fid})$ (calculated for some fiducial value
$M_{\rm fid}$), then an estimator for the halo mass per pixel is
given by
\be
\widehat{M}_i =  \frac{\widehat{\kappa_{\vec{L}_i}}}{\kappa^{\rm
th}_{\vec{L}_i}(M_{\rm fid})}M_{\rm fid},
\label{MassPerPixel}
\ee
with variance
\be
     \langle|\widehat{M}_{i}|^2\rangle =  \frac{\Omega
     N_{\vec{L}_i}}{|\kappa^{\rm th}_{\vec{L}_i}(M_{\rm fid})|^2}M^2_{\rm
     fid}.
\label{MassVarPerPixel}
\ee
From these we can build a minimum-variance estimator,
\be\label{minvarestimator}
\widehat{M} =  \frac{\sum\limits_i
\widehat{M}_i/\langle|\widehat{M}_{i}|^2\rangle}{\sum\limits_i
1/\langle|\widehat{M}_{i}|^2\rangle},
\ee
for the mass as a weighted sum over the pixels.  The variance of
this estimator is, under the null hypothesis,
\bea
\sigma^{-2}_{M} &=&  \sum\limits_i
1/\langle|\widehat{M}_{i}|^2\rangle =  \sum\limits_{\vec L}
\frac{|\kappa^{\rm th}_{\vec L}(M_{\rm fid})|^2}{M^2_{\rm
fid}\Omega N_{\vec L}} \nonumber \\
&\simeq&  \frac{N_z l_{\rm max}^2}{M^2_{\rm
fid}\Omega4\pi}\sum\limits_{\vec L}  |\kappa^{\rm th}_{\vec
L}(M_{\rm fid})|^2|W_{\vec L}|^2 \nonumber \\
&\simeq&  \frac{N_z l_{\rm max}^2}{M^2_{\rm fid}4\pi}\int \frac{d^2\vec{L}}{(2\pi)^2}  |\kappa^{\rm th}_{\vec L}(M_{\rm fid})W_{\vec L}|^2 \nonumber \\
&=& \frac{N_z l_{\rm max}^2 }{M^2_{\rm fid}4\pi} \int d^2 \theta
\left[ \int d^2 \varphi \, W(\vec \varphi) \kappa(|\vec \theta -
\vec \varphi|) \right]^2, \nonumber \\
&&
\label{MassVarTotal}
\eea
where in the second line we substituted 
\be
N_{\vec{L}}=\frac{4\pi}{N_z l_{\rm max}^2}e^{L^2/2L_{\rm
max}^2}\equiv\frac{4\pi}{N_z l_{\rm
max}^2}W_{\vec{L}}^{-1},
\ee
with the exponential included to describe roughly the
transition between those $L$ modes that can be measured and
those that cannot.  In the third line in
Eq.~(\ref{MassVarTotal}) we took the continuum limit $\sum_{\vec
L} \Leftrightarrow \Omega \int d^2\vec L/(2\pi)^2$, and in the
fourth we used Parseval's theorem to switch to real space,
resulting in a convolution of the convergence with a
two-dimensional Gaussian filter,
\be
W(\varphi)=\frac{1}{2\pi \theta_s^2}e^{-\frac{\varphi^2}{2\theta_s^2}},
\ee
with smoothing scale $\theta_s$ given by $\theta_s=\pi/L_{\rm max}$ (we assume here that we can push $L_{\rm max}$ close to $l_{\rm max}$ which is verified numerically). Assuming a spherically symmetric profile and cutting off the integral at $\Lambda$ when the signal becomes negligible, we get
\bea
     \sigma^{-2}_{M} &=&  \frac{N_z l_{\rm max}^2}{2M^2_{\rm
     fid}}\int\limits^\Lambda \, \theta\,
     d\theta \nonumber \\
     &  \times &  \left[
     \int\limits^\Lambda \, \varphi\, d\varphi\, W(\varphi)
     \int\limits_0^{2\pi} \, d\phi\,
     \kappa(\sqrt{\theta^2+\varphi^2+2\theta\varphi\cos\phi})
     \right]^2. \nonumber \\
\label{signaltonoise}
\eea
This equation, together with Eq.~(\ref{l_max}), allows a
straightforward examination of the capabilities of different
telescopes to reconstruct a given lensing source. The focus of
this work is detection of a galaxy cluster, but a similar
formula can be used to estimate the signal-to-noise for
reconstruction of model parameters for other lensing sources, as
we discuss in the Conclusion.

The remaining task is to calculate the convergence profile of a
given cluster.  We model the mass profile of the galaxy cluster
by an NFW profile \cite{NFW},
\be
     \rho(r)=\rho_s{1\over r/r_s(1+r/r_s)^2},
\ee
where the scale radius $r_s$ and normalization $\rho_s$ are
often described by the concentration parameter $c\equiv r_{\rm
vir}/r_s$ and the cluster mass $M\equiv4\pi
r_s^3\rho_s[\ln(1+c)-c/(1+c)]$.  Here, $r_{\rm vir}$ is the
radius within which the enclosed mass $M$ is 200 times the
average mass of the same volume in a critical density universe.

The convergence $\kappa(\theta)$ of the NFW profile is given by
\be
  \kappa(\theta = r/D_L) = \frac{2\rho_{\rm s}r_{\rm
  s}}{\Sigma_{\rm cr}}\,{f(\theta/(r_s/D_L))},
\ee
where
the functional form of the projected mass density is
\be
  f(x) = \begin{cases}
  {1\over x^2 - 1} \left[1 - {2\over\sqrt{x^2 - 1}}\arctan\sqrt{x-1\over x+1}\;\right], & x>1, \cr
  \frac{1}{3}, & x=1, \cr
 {1 \over x^2 - 1} \left[1 - {2\over\sqrt{1 - x^2}}\artanh\sqrt{1-x\over 1+x}\;\right], & x<1. \cr
  \end{cases}
\ee

\begin{figure}
\centering
\includegraphics[width=0.8\linewidth]{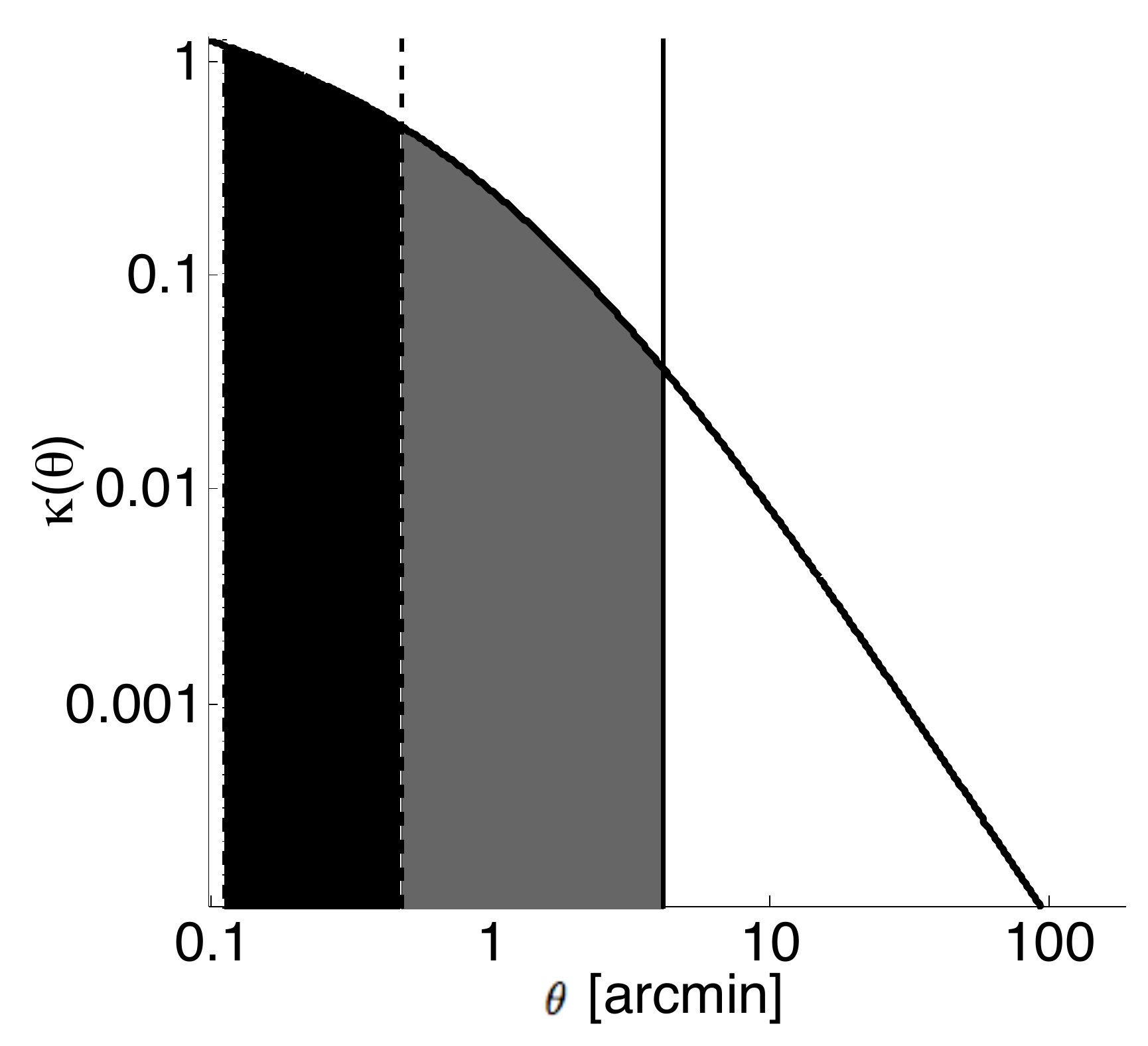}
\caption{The radial convergence of NFW clusters with mass of $M=5\times10^{14}h^{-1}\msun$ and concentration $c=3$ located at redshift $z=1$, where the 21-cm source is at redshift $z=7$. The vertical line marks the virial radii of the clusters where the signal becomes negligible. The two shaded regions show the amount of signal integrated with the ideal resolutions of SKA ({\it light shade}) at $1\,{\rm arcmin}$ and a dark ages observatory ({\it dark shade}) with a $100\,{\rm km}$ baseline at $\sim6\,{\rm arcsec}$.}
\label{fig:ConvergencePerC}
\end{figure}

In Fig.~\ref{fig:ConvergencePerC} we plot the convergence of a
cluster at redshift $z=1$ with mass of
$M_1=5\times10^{14}h^{-1}\msun$ and concentration parameter
$c=3$ for a source at redshift $z=7$. Plugging this in the
integral of Eq.~(\ref{signaltonoise}), we find that the
signal-to-noise is saturated at the virial radius, so that
effectively $\Lambda=r_{\rm vir}$, an order of magnitude above
the corresponding Einstein radius. We also find that the
dependence on the concentration parameter is small. 

\section{Results}
To estimate the signal-to-noise with which this cluster mass can
be reconstructed using the weak-lensing signal of 21-cm
radiation measured by a radio interferometer, we consider the
expected signal from the EOR from SKA (as
described in the previous Section), an upgraded SKA with four
times the collecting area, both with a bandwidth of $1\,{\rm MHz}$, and an ideal experiment (capable of
sustaining $C_l^s>C_l^n$ up to $l_{\rm max}=l_{\rm cover}$) with
the same resolution as SKA ($l_{\rm cover}\gtrsim10^4)$ with a bandwidth small enough to reach the maximum number of independent redshift slices. We also include extremely optimistic limits for an ideal dark ages
observatory with a baseline of $100\,{\rm km}$ (for which $l_{\rm
cover}\sim10^5)$.

\begin{figure}
\centering
\includegraphics[width=0.8\linewidth]{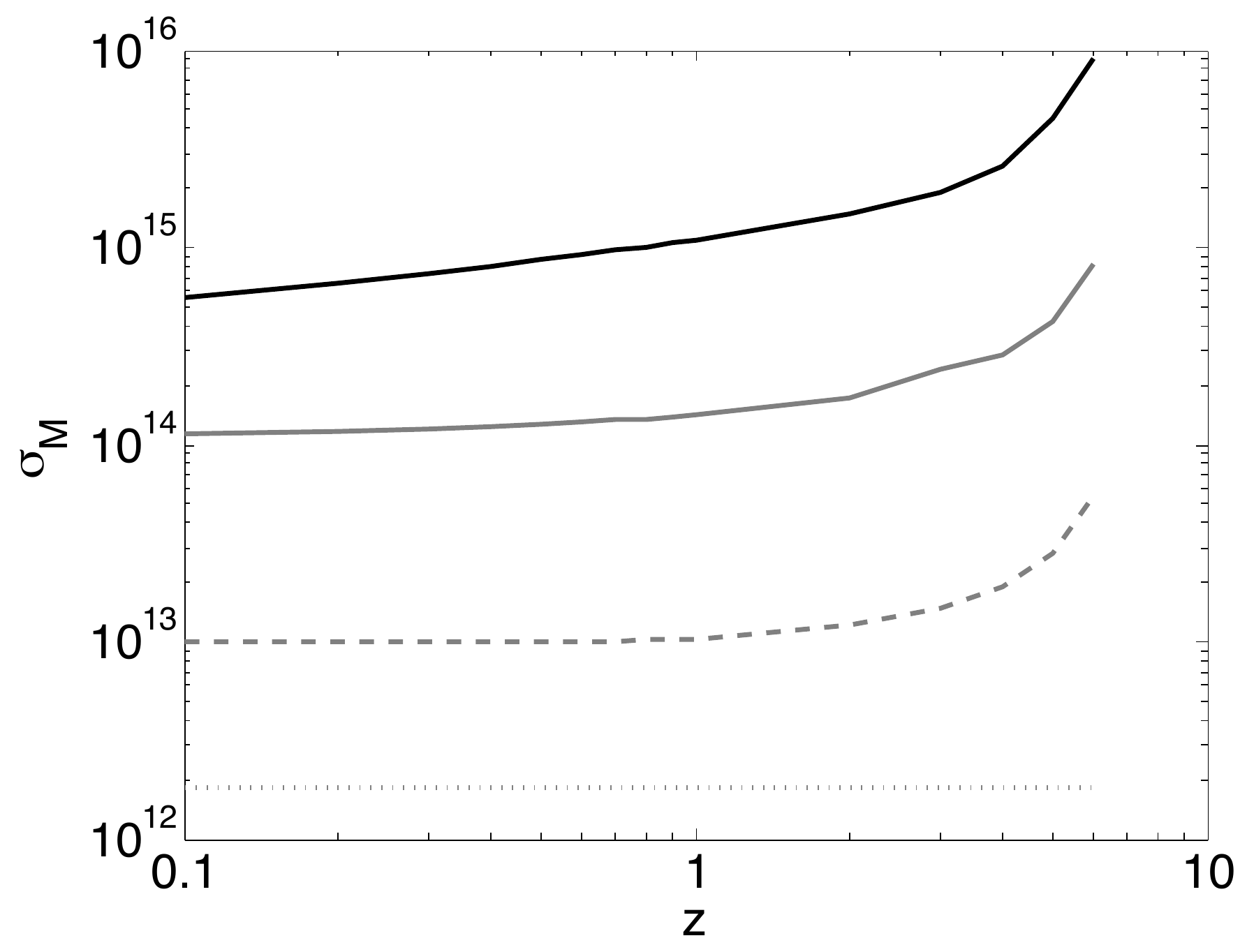}
\caption{The $1\sigma$ limit for weak lensing reconstruction for clusters in different redshifts up to the end of reionization ($z\sim7$) using SKA ({\it \!thick solid line}), SKA with four times the coverage fraction or sixteen times the observing time ({\it \!thin solid line}), both with a bandwidth of $1\,{\rm MHz}$, and for an ideal experiment with SKA resolution ({\it \!dashed line}). We also plot the expected limits for an ideal dark ages observatory with a $100\,{\rm km}$ baseline ({\it \!dotted line}).}
\label{fig:SNPerZLC}
\end{figure}

In Fig.~\ref{fig:SNPerZLC} we plot the smallest detectable
mass as a function of redshift, for different experimental capabilities. We
see that with the current plans for SKA, the weak-lensing
reconstruction of clusters as considered in
Fig.~\ref{fig:ConvergencePerC} is completely beyond reach, unless about an order of magnitude more observation time is dedicated to the patch containing the target galaxy-cluster. Future experiments will narrow this gap and enable the mass measurement of significantly smaller mass halos using 21-cm weak lensing.

These detection prospects involve intricate tradeoffs between the different experimental parameters.  For example, if we increase observation time by 2, we can reach the same $l_{\rm max}$ with half the frequency bandwidth, which will also allow us to use twice the number of z-slices, increasing the S/N by $\sqrt{2}$. Alternatively, we could use this to reach a larger $l_{\rm max}$ (paying a small price for the damping in the signal power spectrum), which also increases the S/N, but this will still be limited by the maximum $l$ set by the baseline of the interferometer. As mentioned in Section IV, one does not always do better by splitting up into smaller bandwidth bins because this increases the noise, yielding a smaller $l_{\rm max}$, which reduces the S/N and might also leave galaxy-cluster scales beyond reach. In addition, as we also discussed in Section IV, there is a lower limit to the frequency bandwidth (which determines the width of the redshift slices) below which the slices become correlated.

Another experimental issue is that as the observed frequency is increased in order to use multiple redshift slices, a larger baseline is needed to cover the same scales. This means that for a given experimental baseline, the contribution of additional slices degrades with their redshift and the $S/N$ grows slower than $1/\sqrt{N_z}$. On the other hand, for high redshift clusters (which approach the sources of the EOR) the increase in lensing signal due to the larger line-of-sight distance traveled is larger than the loss in resolution, and so the $S/N$ grows faster than $1/\sqrt{N_z}$.

\begin{figure}
\centering
\includegraphics[width=0.8\linewidth]{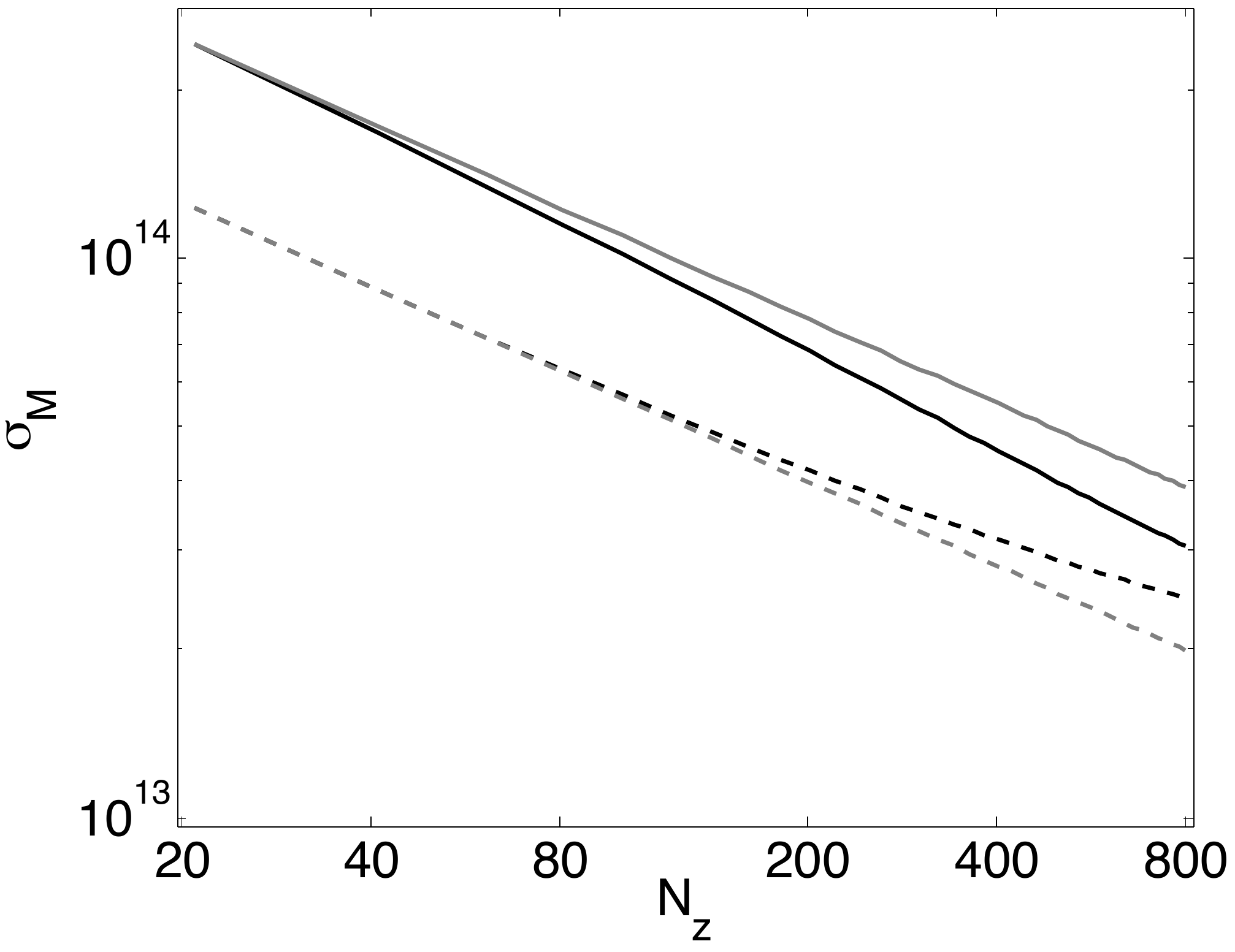}
\caption{The $1\sigma$ limit for weak lensing reconstruction for clusters at two redshifts, $z\!=\!0.5$ ({\it dashed black}) and $z\!=\!5$ ({\it solid black}), as we include more and more redshift slices beyond the end of reionization ($z\sim7$), for SKA with frequency bandwidth $\Delta\nu=0.1{\rm MHz}$) and unlimited observation time. This takes into account the non-trivial source redshift dependence of the different slices. The gray lines are the results under the approximation of a trivial $1/\sqrt{N_z}$ dependence.}
\label{fig:SNPerNz}
\end{figure}

To demonstrate this last point quantitatively\footnote{A full treatment must also take into account the redshift dependence of the 21-cm signal amplitude, which is not trivial, as can be seen for the dark ages in Fig. \ref{fig:CLDarkAges}, but we neglect this for the purpose of this demonstration.}, we plot in Fig.~(\ref{fig:SNPerNz}) the $1\sigma$ detection limits for two cluster redshifts as the number of redshift slices accumulated beyond $z\!=\!7$ is increased. We assume a bandwidth of $\Delta\nu=0.1\,{\rm MHz}$ (yielding roughly $\sim800$ slices in the range $z\!=\!7-13$) and unlimited observation time so that the maximum resolution of SKA is reached. We see that the low redshift cluster gains less improvement in $S/N$ as the number of redshifts is increased, because the smallest observable scale with a given baseline decreases with redshift. We also see that for a high redshift cluster, which is closer to the redshifts of the EOR, the benefit from the inclusion of additional redshift slices more than compensates for this degradation in resolution.

Finally, our estimator, Eq.~\ref{minvarestimator}, can be generalized to include the free parameters of the cluster location, which in this work we assumed were already known from other surveys.

\section{Discussion}

We have constructed a minimum-variance estimator for the mass of
a galaxy cluster using the weak-lensing convergence
reconstruction of 21-cm fluctuations from the dark ages and the EOR.
As has been suggested for CMB lensing reconstruction, possible improvements to the results shown here can stem from
using modified weak-lensing estimators, enhancing sensitivity to small scales as in Ref.~\cite{Maturi} or iteratively approaching a maximum-likelihood estimator as in Ref.~\cite{YooZald}.
Another option which has been discussed for CMB measurements
(e.g. Refs.~\cite{Maturi,YooZald,HuDeDeo}) is to stack reconstructed
images of $N$ different clusters to yield a $\sqrt{N}$
improvement in the signal-to-noise for fitting an overall mass
profile (much like the use of different 21-cm redshift slices of
the same cluster in Eq.~(\ref{volumenoise})). 

In comparison to other sources for weak-lensing measurements,
such as galaxy shapes \cite{Blandford,Bacon}, or
CMB fluctuations \cite{Seljak} (detected recently \cite{Smith}),
the potential in 21-cm measurements is far greater. Even deep
galaxy surveys will be limited to lower redshifts with small sky
coverage and will not exceed arcminute resolutions. CMB lensing
reconstruction at small scales is difficult because of the absence of power due to Silk
damping at these scales and due to the ambiguity caused by the presence of a similar signature from the kinetic
Sunyaev-Zeldovich effect on the CMB photons passing through the
moving clusters \cite{YooZald,HuDeDeo}. 
Even with ideal experiments at arcminute resolutions, the corresponding minimum galaxy-cluster mass detectable according to Eq.~(\ref{signaltonoise}) with a CMB experiment is $M\sim10^{15}h^{-1}\msun$. An average mass profile of $M\sim10^{14}h^{-1}\msun$ can be measured with reasonable signal to noise by stacking thousands of galaxies, but the detailed reconstruction of an individual cluster is beyond the reach of these
alternative methods.

Finally, an issue which we have neglected in this work is the influence
of non-gaussianities which would have to be taken into account
to yield accurate predictions. Particularly important is the
effect of nonlinear structure \cite{LuPen,LuPenDore}, which is
relevant during the epoch of reionization, when certain patches
of the intergalactic medium become substantially ionized well before its
end. In order to properly account for these effects, the details of the reionization process will have to be uncovered.
However, under the assumption that these features appear in
higher resolutions than that of our considered interferometers
\cite{ZahnZald,MetcalfWhite}, there will be no connected
four-point function contribution to the variance of the quadratic lensing estimator used here, and this treatment remains valid.

\section{Conclusion}

When 21-cm fluctuations become accessible to observations at
small angular scales, the application of weak-lensing
reconstruction methods will open the door to unprecedented
precision measurements of local structure. We have found here
that galaxy clusters can be detected by lensing reconstruction
with futuristic experiments measuring CMB absorption by neutral
hydrogen during the dark ages $30 \lesssim z \lesssim 200$ down
to halo masses of order $M\gtrsim10^{12} h^{-1}\,M_\odot$. Next-generation
interferometers measuring emission from 
hot neutral-hydrogen gas during the epoch of reionization
$7\lesssim z \lesssim 13$ will be limited to $M\gtrsim10^{15}\,
h^{-1}\,M_\odot$ with the currently planned specifications.  If,
however,  the collecting area or the observation time can be
increased, they may be able to push the limit down to
$M\gtrsim10^{13}\, h^{-1}\, M_\odot$ (in particular, this can be reached by increasing the observation time dedicated to the target patch alone by two orders of magnitude).
While these are challenging goals that remain unattainable in the near future,
the potential achievements discussed here provide more
motivation to invest in alleviating the experimental
limitations.

Meanwhile, we can use the prescription described here to
construct estimators for model parameters of other structures,
from standard isothermal spheres and voids to more exotic
structures such as cosmic textures \cite{Turok,TurokSpergel} or
overdensities created by pre-inflationary particles
\cite{Fialitzkov,Rathaus}. In future work we shall address the
weak-lensing detectability, with 21-cm weak lensing, of voids or
textures that might be responsible for the cold spot in WMAP CMB
data \cite{DasSpergel,BeNaSunny}.

\acknowledgments
We thank the anonymous referee for excellent comments and suggestions which helped improve the quality of this paper. EDK was supported by the National Science Foundation under Grant Number PHY-0969020 and by the Texas Cosmology Center. This work was supported at Johns Hopkins by DoE SC-0008108 and NASA NNX12AE86G. EDK would like to thank the hospitality of the Department of Physics and Astronomy at Johns Hopkins, where this work was partially carried out.  

\appendix
\section{Power spectrum of 21-cm fluctuations}

We review the angular power spectrum of intensity
fluctuations in the 21-cm signal induced by large-scale density
inhomogeneities during the dark ages and the EOR, when redshift distortions are neglected.  Following
Refs.~\cite{Zaldarriaga,Furlanetto}, we start in three dimensions and
first define the dimensionless fractional perturbation
$\delta_{HI}({\bf x}) \equiv [\dtb({\bf x}) - \bdtb]/\bdtb$ of
the brightness temperature (where $\bdtb$ is its mean), whose power
spectrum in Fourier space is given by $\langle
\delta_{HI}(\bk_1) \, \delta_{HI}(\bk_2) \rangle \equiv (2
\pi)^3 \delta^3(\bk_1 + \bk_2) P(k_1,z)$ with its dimensionless
counterpart $\Delta^2(k,z) = (k^3/2 \pi^2) P(k,z)$.
We move to two dimensions by integrating along the line of sight,
\begin{equation}
\delta_{HI}(\bhn,\nu) = \int \deriv r \, W(r, r_{\nu}) \, \delta_{HI}(\bhn,r),
\label{eq:windowfunction}
\end{equation}
with a projection window function $W(r,r_{\nu})$ that peaks at
the radial distance $r_{\nu}$ to the desired frequency $\nu$
within a width $\delta r$ corresponding to an observational
bandwidth $\Delta\nu$. To expand the brightness temperature in spherical harmonics, we use the planar wave expansion $e^{i\bk \cdot {\bf x}} = \sum_{lm} 4 \pi i^l j_l(kr) Y_{lm}^*(\bhk) Y_{lm}(\bhn)$ (where $j_l(x)$ is the spherical Bessel function of order $l$), and
\bea
a_{lm}(\nu) & = & 4 \pi i^l \int \frac{\deriv^3 k}{(2 \pi)^3} \delta_{HI}({\bf k},\nu)\alpha_l(k,\nu) Y_{lm}^*(\bhk) \nonumber \\
\alpha_l(k,\nu)&=&\bdtb(\nu)\int \deriv r \, W(r, r_{\nu}) \, j_l(kr)~.
\eea
The brightness temperature power spectrum of 21-cm fluctuations, defined as $\langle{ a_{lm}(\nu) a^*_{l' m'}(\nu')}\rangle \equiv \delta_{l l'} \delta_{m m'} C_{l}(\nu,\nu')$, is then given by
\be
C_l(\nu, \nu')= 4\pi\int \deriv k \frac{\Delta^2(k,z)}{k}\alpha_l(k,\nu)\alpha_l(k,\nu').
\label{eq:21PowerSpectrum}
\ee
Under the approximation $\Delta^2(k,z)\approx\Delta^2(l/r_{\nu},z)$, for a pure power-law power spectrum and for large angular scales $l \delta r/r \ll 1$, the line of sight integration is approximately a delta function and we get
\begin{equation}
\frac{l^2 C_l(\nu, \nu)}{2 \pi} \propto \bdtb^2(\nu) \Delta^2(l/r_{\nu},z).
\label{eq:21PowerSpectrumLargeScales}
\end{equation}
For the purposes of reconstructing galaxy-cluster profiles, we are mostly interested in the smallest observable scales, which
satisfy the limit $l \delta r/r \gg 1$. Applying the Limber
approximation \cite{Limber,Peebles} in Fourier space
\cite{Kaiser} yields 
\begin{equation}
\frac{l^2 C_l(\nu, \nu)}{2 \pi} \propto \bdtb^2(\nu) \Delta^2(l/r_{\nu},z) \frac{r_{\nu}}{l \, \delta r},
\label{eq:21PowerSpectrumSmallScales}
\end{equation}
where we see the suppression of small-angle fluctuations
resulting from averaging out of modes $k \gtrsim 1/\delta r$ in
Eq.~(\ref{eq:windowfunction}).


\begin{thebibliography}{99}

\bibitem{Furlanetto} 
  S.~Furlanetto, S.~P.~Oh and F.~Briggs,
  Phys.\ Rept.\  {\bf 433}, 181 (2006)
  [astro-ph/0608032].

\bibitem{Pritchard:2011xb} 
  J.~R.~Pritchard and A.~Loeb,
  Rept.\ Prog.\ Phys.\  {\bf 75}, 086901 (2012)
  [arXiv:1109.6012 [astro-ph.CO]].

\bibitem{LOFARSKA}
  J.\ D.\ Bowman and A.\ E.\ E.\ Rogers, Nature {\bf 468}, 796
  (2010); G.\ Paciga {\it et al.} (GMRT-EoR Collaboration),
  Mon.\ Not.\ R.\ Astron.\ Soc.\ {\bf 413}, 1174 (2011); {\tt
  www.lofar.org}; {\tt www.mwatelescope.org}; {\tt
  web.phys.cmu.edu/~past}; {\tt www.phys.unm.edu/~lwa}; {\tt
  astro.berkeley.edu/~dbacker/eor}; {\tt
  www.haystack.mit.edu/ast/arrays/Edges}; {\tt
  www.skatelescope.org}.

\bibitem{Lunar} 
  S.~Jester and H.~Falcke, New Astron.\ Rev.\ \ {\bf 53}, 1  (2009) [arXiv:0902.0493 [astro-ph.CO]];
T. J.~W. Lazio, J.~Burns, D.~Jones, J.~Kasper, S.~Neff, R.~MacDowall, K.~Weiler, and
DALI/ROLSS Team, Bulletin of the American Astronomical Society
41, 344 (2009);
J.~O.~Burns, T.~J.~W.~Lazio and W.~Bottke,
  arXiv:1209.2233 [astro-ph.CO].

\bibitem{Zaldarriaga} 
  M.~Zaldarriaga, S.~R.~Furlanetto and L.~Hernquist,
  Astrophys.\ J.\  {\bf 608}, 622 (2004)
  [astro-ph/0311514].
  
\bibitem{LoebZald}
 A.~Loeb and M.~Zaldarriaga,
 Phys.\ Rev.\ Lett.\  {\bf 92}, 211301 (2004)
 [arXiv:astro-ph/0312134].
  
 \bibitem{CAMB_sources} 
  A.~Lewis and A.~Challinor,
  Phys.\ Rev.\ D {\bf 76}, 083005 (2007)
  [astro-ph/0702600 [ASTRO-PH]].

\bibitem{ZahnZald} 
  O.~Zahn and M.~Zaldarriaga,
  Astrophys.\ J.\  {\bf 653}, 922 (2006)
  [astro-ph/0511547].

\bibitem{Metcalf} 
  R.~B.~Metcalf and S.~D.~M.~White,
  Mon.\ Not.\ Roy.\ Astron.\ Soc.\  {\bf 394}, 704 (2009)
  [arXiv:0801.2571 [astro-ph]].
  
\bibitem{MetcalfWhite} 
R.~B.~Metcalf and S.~D.~M.~White,
  Mon.\ Not.\ Roy.\ Astron.\ Soc.\  {\bf 381}, 447 (2007)
  [arXiv:0611862 [astro-ph]].

\bibitem{Hilbert} 
  S.~Hilbert, R.~B.~Metcalf and S.~D.~M.~White,
  Mon.\ Not.\ Roy.\ Astron.\ Soc.\  {\bf 382}, 1494 (2007)
  [arXiv:0706.0849 [astro-ph]].

\bibitem{LuPen} 
  T.~Lu and U.~-L.~Pen,
  [arXiv:0710.1108 [astro-ph]].

\bibitem{Sigurdson:2005cp} 
  K.~Sigurdson and A.~Cooray,
  Phys.\ Rev.\ Lett.\  {\bf 95}, 211303 (2005)
  [astro-ph/0502549].

\bibitem{Book:2011dz} 
  L.~Book, M.~Kamionkowski and F.~Schmidt,
  Phys.\ Rev.\ Lett.\  {\bf 108}, 211301 (2012)
  [arXiv:1112.0567 [astro-ph.CO]].

\bibitem{Seljak}
  M.~Zaldarriaga, U.~Seljak,
  Phys.\ Rev.\  {\bf D59}, 123507 (1999)
  [astro-ph/9810257];
  U.~Seljak and M.~Zaldarriaga,
  Phys.\ Rev.\ Lett.\  {\bf 82}, 2636 (1999)
  [astro-ph/9810092];
  W.~Hu,
  Phys.\ Rev.\  {\bf D64}, 083005 (2001)
  [astro-ph/0105117];
  Astrophys.\ J.\  {\bf 557}, L79 (2001)
  [astro-ph/0105424];
  W.~Hu and T.~Okamoto,
  Astrophys.\ J.\  {\bf 574}, 566 (2002)
  [astro-ph/0111606];
  M.~H.~Kesden, A.~Cooray and M.~Kamionkowski,
  Phys.\ Rev.\  {\bf D67}, 123507 (2003)
  [astro-ph/0302536];
  A.~Lewis and A.~Challinor,
  Phys.\ Rept.\  {\bf 429}, 1 (2006)
  [astro-ph/0601594];
  S.~Dodelson, F.~Schmidt and A.~Vallinotto,
  Phys.\ Rev.\ D\ {\bf 78}, 043508  (2008)
  [arXiv:0806.0331 [astro-ph]];
  A.~Cooray, M.~Kamionkowski and R.~R.~Caldwell,
  Phys.\ Rev.\ D {\bf 71}, 123527 (2005)
  [astro-ph/0503002].

\bibitem{Okamoto} 
  T.~Okamoto and W.~Hu,
  Phys.\ Rev.\ D {\bf 67}, 083002 (2003)
  [astro-ph/0301031].
  
\bibitem{HuDeDeo} 
  W.~Hu, S.~DeDeo and C.~Vale,
  New J.\ Phys.\  {\bf 9}, 441 (2007)
  [astro-ph/0701276].

\bibitem{Silk:1967kq} 
  J.~Silk,
  Astrophys.\ J.\  {\bf 151}, 459 (1968).
  
  \bibitem{Albrecht} 
  A.~Albrecht {\it et al.},
  ArXiv Astrophysics e-prints (2006),
  [astro-ph/0609591].

\bibitem{field59}
G.~B.~Field,
Astrophys.\ J.\  {\bf 129}, 525 (1959).

\bibitem{Limber} 
  D.~N.~Limber, Astrophys.\ J.\  {\bf 117}, 134 (1953);

\bibitem{Peebles}
  P.~J.~E.~Peebles,
  {\it The Large-Scale Structure of the Universe} (Princeton
  University Press, Princeton, 1980).

\bibitem{Kaiser} 
  N.~Kaiser, Astrophys.\ J.\  {\bf 388}, 272 (1992).

\bibitem{NFW} 
  J.~F.~Navarro, C.~S.~Frenk and S.~D.~M.~White,
  Astrophys.\ J.\  {\bf 490}, 493 (1997)
  [astro-ph/9611107].
  
\bibitem{Maturi} 
  M.~Maturi, M.~Bartelmann, M.~Meneghetti and L.~Moscardini,
  Astron.\ Astrophys.\  {\bf 436}, 37 (2005)
  [astro-ph/0408064].

\bibitem{YooZald} 
  J.~Yoo and M.~Zaldarriaga,
  Phys.\ Rev.\ D {\bf 78}, 083002 (2008)
  [arXiv:0805.2155 [astro-ph]].

\bibitem{Blandford}
 R.~D.~Blandford {\it et al.},
 Mon.\ Not.\ Roy.\ Astron.\ Soc.\  {\bf 251}, 600 (1991);
  J.\ Miralda-Escud\'e, Astrophys.\ J.\ {\bf 380}, 1 (1991);
  N.\ Kaiser, Astrophys.\ J.\ {\bf 388}, 272 (1992);
  M.\ Bartelmann and P.\ Schneider, Astron.\ Astrophys. {\bf
  259}, 413 (1992);
  M.~Kamionkowski, A.~Babul, C.~M.~Cress and A.~Refregier,
  Mon.\ Not.\ Roy.\ Astron.\ Soc.\  {\bf 301}, 1064 (1998)
  [astro-ph/9712030];
  M.~Bartelmann and P.~Schneider,
  Phys.\ Rept.\  {\bf 340}, 291 (2001)
  [astro-ph/9912508].

\bibitem{Bacon}
  D.~J.~Bacon, A.~R.~Refregier and R.~S.~Ellis,
  Mon.\ Not.\ Roy.\ Astron.\ Soc.\  {\bf 318}, 625 (2000)
  [astro-ph/0003008];
  N.~Kaiser, G.~Wilson and G.~A.~Luppino,
  Astrophys.\ J.\ {\bf 556}, 601 (2001) [astro-ph/0003338];
  D.~M.~Wittman {\it et al.},
  Nature {\bf 405}, 143 (2000)
  [astro-ph/0003014];
  L.~van Waerbeke {\it et al.},
  Astron.\ Astrophys.\  {\bf 358}, 30 (2000)
  [astro-ph/0002500].

\bibitem{Smith}
 K.~M.~Smith, O.~Zahn and O.~Dore,
 Phys.\ Rev.\  D {\bf 76}, 043510 (2007)
 [arXiv:0705.3980 [astro-ph]];
 C.~M.~Hirata {\it et al.},
 Phys.\ Rev.\  D {\bf 78}, 043520 (2008)
 [arXiv:0801.0644 [astro-ph]];
 S.~Das {\it et al.} (ACT Collaboration),
 Phys.\ Rev.\ Lett.\  {\bf 107}, 021301 (2011)
 [arXiv:1103.2124 [astro-ph.CO]].
 
  \bibitem{LuPenDore} 
  T.~Lu, U.~-L.~Pen and O.~Dore,
  Phys.\ Rev.\ D {\bf 81}, 123015 (2010)
  [arXiv:0905.0499 [astro-ph.CO]].
   
\bibitem{Turok} 
  N.~Turok,
  Phys.\ Rev.\ Lett.\  {\bf 63}, 2625 (1989).
  
\bibitem{TurokSpergel} 
  N.~Turok and D.~Spergel,
  Phys.\ Rev.\ Lett.\  {\bf 64}, 2736 (1990).
  
\bibitem{Fialitzkov} 
  A.~Fialkov, N.~Itzhaki and E.~D.~Kovetz,
  JCAP {\bf 1002}, 004 (2010)
  [arXiv:0911.2100 [astro-ph.CO]].
  
\bibitem{Rathaus} 
  B.~Rathaus and N.~Itzhaki,
  JCAP {\bf 1205}, 006 (2012)
  [arXiv:1202.5178 [astro-ph.CO]].
  
\bibitem{DasSpergel} 
  S.~Das and D.~N.~Spergel,
  Phys.\ Rev.\ D {\bf 79}, 043007 (2009)
  [arXiv:0809.4704 [astro-ph]].

\bibitem{BeNaSunny} 
  B.~Rathaus, A.~Fialkov and N.~Itzhaki,
  JCAP {\bf 1106}, 033 (2011)
  [arXiv:1105.2940 [astro-ph.CO]].

\end{thebibliography}
\end{document}